# Indefinite and Bidirectional Near Infrared Nanocrystal Photoswitching

*Changhwan Lee[1], Emma Z. Xu[1], Kevin W. C. Kwock[2], Ayelet Teitelboim[3], Yawei Liu[3,4], Natalie Fardian-Melamed[1], Cassio C. S. Pedroso[3], Hye Sun Park[5], Jongwoo Kim[6], Stefanie D. Pritzl[7,8], Sang Hwan Nam[6], Theobald Lohmueller[7], Peter Ercius[3], Yung Doug Suh[9],\*, Bruce E Cohen[3,10],\*, Emory M Chan[3],\*, P. James Schuck[1],\**

[1]Department of Mechanical Engineering, Columbia University, New York, New York 10027, United States
[2]Department of Electrical Engineering, Columbia University, New York, New York 10027, United States
[3]The Molecular Foundry, Lawrence Berkeley National Laboratory, Berkeley, CA, USA.
[4]State Key Laboratory of Rare Earth Resource Utilization, Changchun Institute of Applied Chemistry, Chinese Academy of Sciences, Changchun, China
[5]Research Center for Bioconvergence Analysis, Korea Basic Science Institute (KBSI), Cheongju, South Korea
[6]Laboratory for Advanced Molecular Probing (LAMP), Korea Research Institute of Chemical Technology (KRICT), Daejeon, South Korea
[7]Chair for Photonics and Optoelectronics, Nano-Institute Munich, Ludwig-Maximilians Universität München, Germany
[8]SCMB, Debye Institute, Leonard S. Ornstein Laboratory, Utrecht University, Princetonplein 1, 3584 CC Utrecht, The Netherlands
[9]Department of Chemistry, Ulsan National Institute of Science and Technology (UNIST), Ulsan, South Korea
[10]Division of Molecular Biophysics and Integrated Bioimaging, Lawrence Berkeley National Laboratory, Berkeley, CA, USA

*[\*]p.j.schuck@columbia.edu; emchan@lbl.gov; becohen@lbl.gov; ydsuh@unist.ac.kr*

**Materials whose luminescence can be switched by optical stimulation drive technologies ranging from superresolution imaging[1-4], nanophotonics[5], and optical data storage[6-8], to targeted pharmacology, optogenetics, and chemical reactivity[9]. These photoswitchable probes, including organic fluorophores and proteins, are prone to photodegradation, and often require phototoxic doses of ultraviolet (UV) or visible light. Colloidal inorganic nanoparticles have significant stability advantages over existing photoswitchable materials, but the ability to switch emission bidirectionally, particularly with NIR light, has not been reported with nanoparticles. Here, we present 2-way, near-infrared (NIR) photoswitching of avalanching nanoparticles (ANPs), showing full optical control of upconverted emission using phototriggers in the NIR-I and NIR-II spectral regions useful for subsurface imaging. Employing single-step photodarkening[10-13] and photobrightening[12,14-18], we demonstrate indefinite photoswitching of individual nanoparticles (>1000 cycles over 7 h) in ambient or aqueous conditions without measurable photodegradation. Critical steps of the photoswitching mechanism are elucidated by modeling and by measuring the photon avalanche properties of single ANPs in both bright and dark states. Unlimited, reversible photoswitching of ANPs enables indefinitely rewritable 2D and 3D multi-level optical patterning of ANPs, as well as optical nanoscopy with sub-Å localization superresolution that allows us to distinguish individual ANPs within tightly packed clusters.**

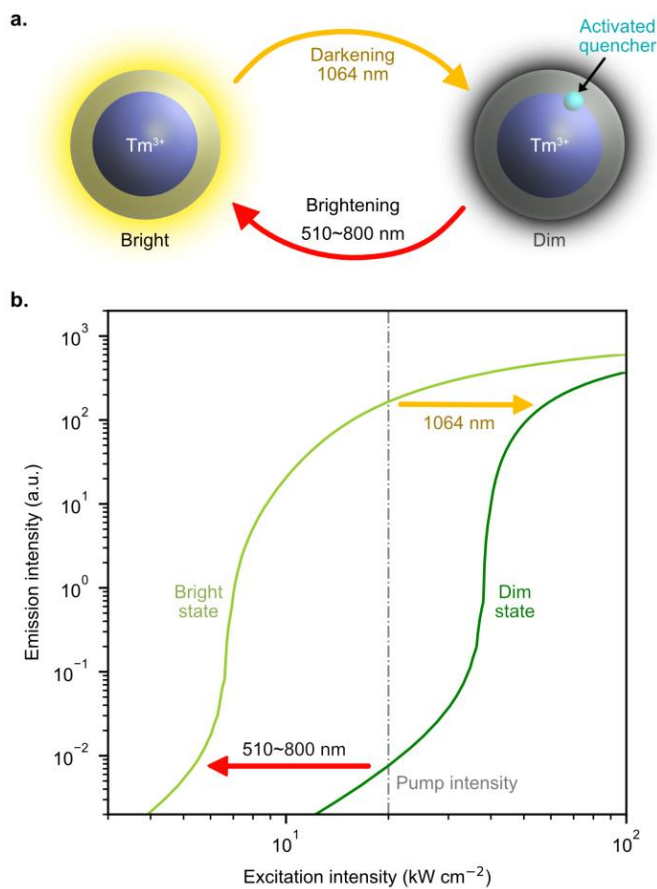

**Fig. 1| Photoswitchable photon avalanching nanoparticles**. **a,** Schematic highlighting the fully reversible NIR photoswitching response of ANPs. Emission is photodarkened using 1064 nm light and photobrightened using wavelengths between 510 and 800 nm. **b,** Optical manipulation of the photon avalanching threshold in ANPs enables photoswitching behavior. The plots are derived from the DRE fits of experimental data in Fig. 3a.

Upconverting nanoparticles (UCNPs) are lanthanide ion ($Ln^{3+}$)-based phosphors that efficiently convert NIR light to higher energies in the NIR, visible, or UV regions[19-22]. Unlike organic[4,23-29], protein[26,30,31], or hybrid organic-inorganic luminescent probes[32], UCNPs do not measurably photobleach, even with extended single-particle excitation under ambient conditions[20,33-35], or within microlasers under high pump powers[36,37]. While this exceptional photostability suggests $Ln^{3+}$-based UCNPs cannot be modulated by light, certain $Ln^{3+}$-based bulk materials have been reported to be susceptible to photodarkening or brightening[11,38-41]. Based on observations of photodarkening in $Tm^{3+}$-doped fibres[11,39,41], as well as studies of colour centers and charge traps in UCNPs[16,17,42,43], we sought to determine if $Tm^{3+}$-doped avalanching nanoparticles (ANPs) could be modulated by light in the same manner. ANPs are UCNPs with the steepest nonlinear emission response of any nanoscale material[22,44,45], enabled by the photon avalanching upconversion mechanism. Single-ANP characterization has shown that relatively minor variations in shell thickness lead to sharp changes in avalanche threshold[46]. This suggests that light-induced charge or energy transfer within the nanocrystal might also shift the PA threshold, magnifying the influence of a minute density of trap states into major emission differences (Fig. 1, 3a).

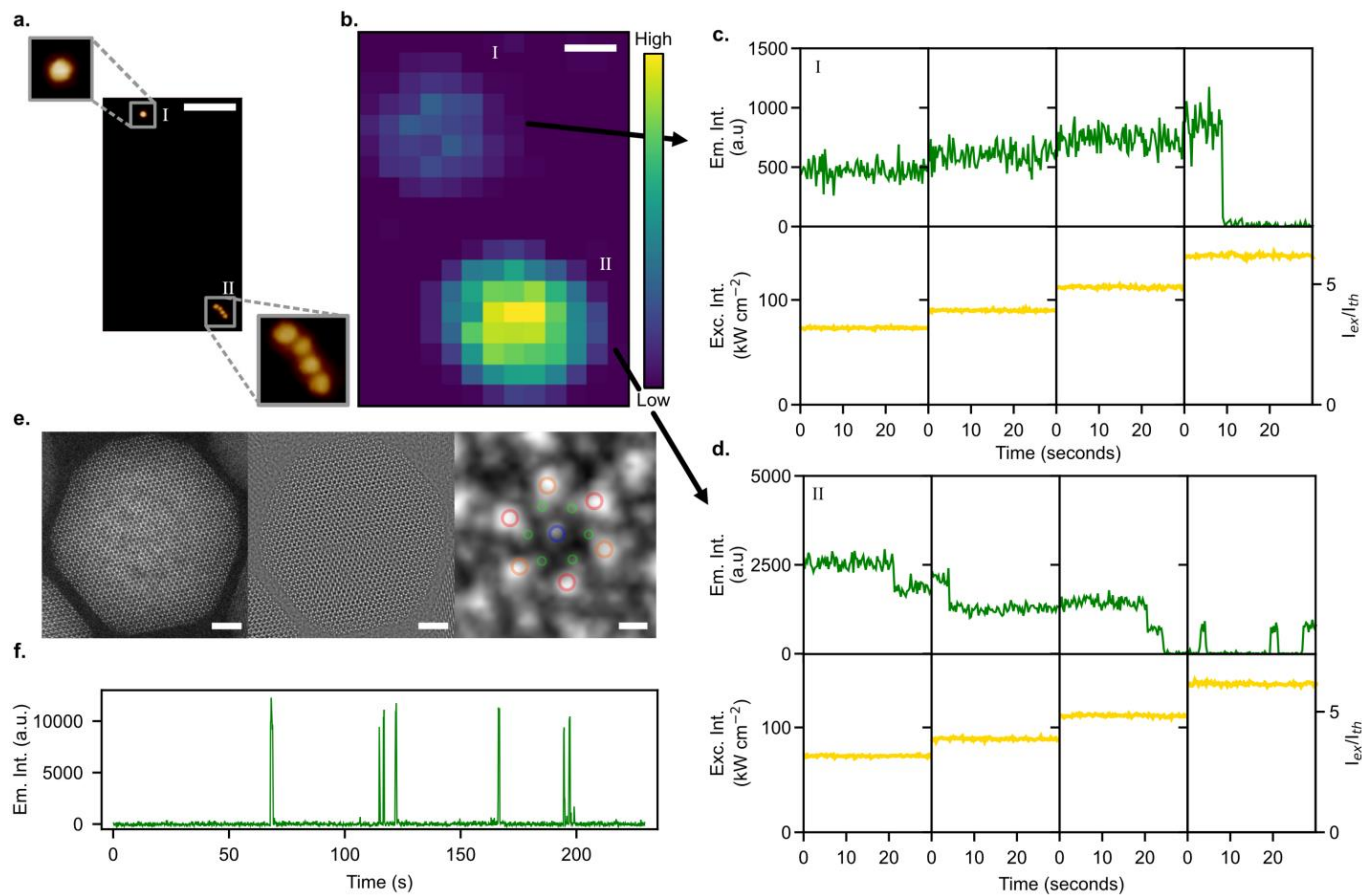

**Fig. 2| Photodarkening and photoblinking in ANPs.** (**a**) AFM and (**b**) confocal scanning images of a single and a cluster of 4 ANPs (NaYF$_4$: 8% Tm$^{3+}$@NaY$_{0.8}$Gd$_{0.2}$F$_4$, 10nm core/4 nm shell). Scale bars are 250 nm. Magnified AFM images of the ANPs are shown in the top left (single) and bottom right (4 singles) panels in **a**. Color bar in **b**: normalized luminescence intensity. Luminescence and excitation intensity $I_{ex}$ time-traces of the single (**c**) and four-ANP cluster (**d**) in **a** under 1064 nm excitation at increasing intensities. $I_{th}$, ANP avalanching threshold intensity. **e,** High-angle annular dark-field scanning transmission electron microscopy (HAADF-STEM, left), differential phase contrast (DPC)-STEM (middle) and magnified NaYF$_4$ unit cell (right) images of an ANP. Scale bars are 4 nm (left and middle) and 2 Å (right). Red, orange, green, and blue markers are 50% Na/50% Y, Y, F, and Na ions, respectively. **f**, Time trace showing blinking luminescence from a single 8% Tm$^{3+}$ 17/6 nm core/shell nanocrystal at $I_{ex}$ = 164 kW cm$^{-2}$.

To determine if ANPs are capable of photoswitching, we characterized single, core/shell NaYF$_4$ ANPs with 8% Tm$^{3+}$ under 1064-nm excitation, at a range of power densities (Fig. 2). By atomic resolution STEM (Fig. 2e), these ANPs are pure β-NaYF$_4$ without any observable extended defects or a clear core/shell interface (note, STEM cannot observe point defects). Above the avalanching threshold intensity $I_{th}$, the ANPs exhibit luminescence at 800 nm. As the pump intensity is increased well above $I_{th}$ (Fig. 2), we observe that single ANPs darken or blink, exhibiting discontinuous jumps in luminescence (Fig. 2c,d) within the time-resolution of our measurement (see Fig. S3 and Methods). The observed single-ANP *photoblinking* is evidence that the photodarkening process – to date viewed as undesirable but unavoidable in bulk materials – is not permanent and can be reversed. Photoblinking statistics from single ANPs show that the blinking is intensity dependent and can be tuned to match desired parameters for some single-molecule localization microscopy (SMLM) applications such as stochastic optical reconstruction microscopy (STORM) imaging (Fig 2f, Fig. S7, S8 and S9)[4].

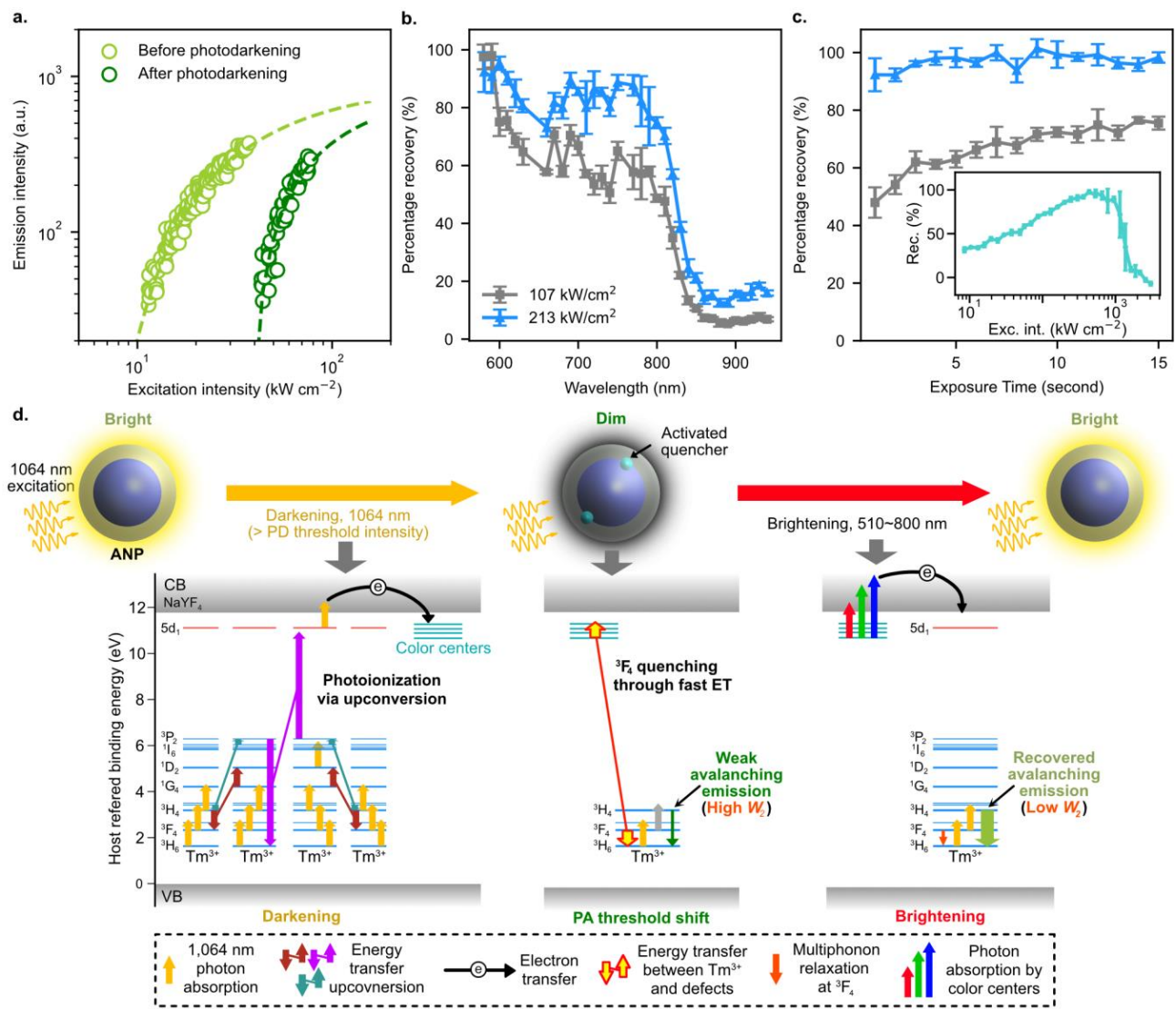

**Fig. 3 | ANP Photobrightening and photodarkening. a,** 800-nm emission intensity versus 1064-nm excitation intensity of a single 8% $Tm^{3+}$ 17.3/5.6 nm core/shell ANP before and after photodarkening. Photoswitching between bright and dim states in ANPs is induced by manipulation of the PA threshold, using 1064 nm light to darken and wavelengths ≤800 nm to brighten. **b,** Percentage photobrightening recovery of a photodarkened 8% $Tm^{3+}$ 10.2/4.0 nm core/shell ANP ensemble film sample versus irradiation wavelength. **c,** Percentage recovery versus exposure time and irradiation intensity (inset). Percentage recovery is defined as the ratio of the recovered luminescence intensity after irradiation with <800nm light to the reduced luminesce intensity after initial photodarkening. Irradiation wavelength is 700 nm in **c**. The photobrightening irradiation intensities in **b** and **c** are shown in the legend in **b**. The photodarkening conditions are 1064 nm irradiation at 261 kW $cm^{-2}$ for 10 s. Error bars are standard deviations of data points from 4 separate measurements. **d,** Potential mechanistic pathways for photodarkening and photobrightening in ANPs. CB: conduction band, VB: valence band, $W_2$: $^3F_4$ relaxation rate.

Additional measurements in ensemble films of 4% and 8% $Tm^{3+}$ nanocrystals suggest that photon avalanching (PA) plays a central role in the photodarkening. UCNPs doped with ~4% $Tm^{3+}$, whose nonlinear emission is sub-avalanching (energy looping[47]), do not show the same photodarkening or blinking behavior, at least for excitation intensities up to 2.3 MW cm$^{-2}$ (Fig. S10). When doping is increased to 8% Tm$^{3+}$, the NPs exhibit pronounced PA and noticeable photodarkening at excitation intensities ≥143 kW cm$^{-2}$ (~6× $I_{th}$, Fig. S11). We further observe that the rate of photodarkening accelerates as pump intensity increases (Fig. S11). The ANP photodarkening and blinking are observed in particles with both thin (2.6 nm) and thick (8.5 nm) inert shells, as well as in ANPs in which passivating oleic acid ligands have been removed, suggesting that surface quenching does not play a major role. We also note that this dependence of photodarkening on Ln$^{3+}$ content and pump intensity is consistent with the reported behaviours of Tm$^{3+}$-doped fibres under intense 1064 or 1120 nm excitation[11,39,41].

To better understand the origins of the observed ANP behaviors, we examined the nonlinear emission of single ANPs in both bright and photodarkened states. Photodarkened nanocrystals continue to exhibit PA emission, but with avalanching threshold intensities $I_{th}$ shifted to ~5-fold higher pump intensities (Fig. 3a). Because of the steeply nonlinear PA process, the luminescence intensity at the same 1064 nm pump intensity is reduced by >4 orders of magnitude following this 5-fold increase in $I_{th}$ (Fig. S12)[22]. In this case, a saturating pumping intensity at first leads to bright ANP luminescence, but then shifts to weak pre-avalanche luminescence upon the attendant $I_{th}$ shift (Figs. 3a, S12 and S13).

Next, motivated by the photoblinking, we studied the potential for controllable luminescence recovery. In this experiment, we first photodarkened a region of an ANP film with intense 1064 nm pumping, then excited the region with a laser tunable from 510 nm to 940 nm (Methods). Plots of luminescence recovery *versus* photobrightening wavelength (Figs. 3b, S14) show between ~10 - 100% recovery in ANP films, depending on illumination wavelength. Increasing the irradiation intensity also enhances photobrightening (inset in Fig. 3c). The log-log slope of the recovery *versus* irradiation plot is sublinear, in sharp contrast to the avalanching emission of these ANPs ($s > 20$). At much more intense photobrightening intensities (>500 kW cm$^{-2}$), additional photodarkening is ultimately induced (inset in Fig. 3c), which is also recoverable. These photobrightening parameter studies (Fig. 3b,c) demonstrate a wide range of control over photoswitching probabilities in the ANPs.

In addition to showing that photoswitching is possible, the photodarkening and photobrightening measurements provide important insights into the mechanisms at play. Specifically, in addition to the shift in threshold intensity described above (Figs. 1b and 3a), any mechanism for ANP photoswitching needs to account for the following key observations: 1) both the bright and dark states are stable and show similar steeply non-linear emission; 2) the darkening is fully reversible, with no measurable long-term degradation in >1000 cycles (see below); 3) the ability to photodarken is unrelated to factors affecting surface quenching (*i.e.*, surface ligands, water/O$_2$, or shell thickness); and 4) the brightening step is possible with a wavelength range much broader than typical for Ln$^{3+}$ excitation. To better explain this, we fit the experimental power-dependent emission curves for darkened and undarkened ANPs to a rate equation model that we previously developed to describe the population balance and the radiative and nonradiative relaxation processes in ANPs[22] (Fig. 3a, dashed lines). Fits to the model reveal that, compared to undarkened ANPs, the darkened ANP have a 5.3-fold faster overall relaxation rate ($W_2$) from the $^3F_4$ first excited state of Tm$^{3+}$ (SI). This suggests an added, faster, loss of energy from $^3F_4$.

Taken together, our models and observations are consistent with a photodarkening scheme in which extended ANP photoexcitation populates high-lying Tm$^{3+}$ excited states (*e.g.*, $^1I_6$ and 5d$^1$4f$^{11}$ states[48]; see

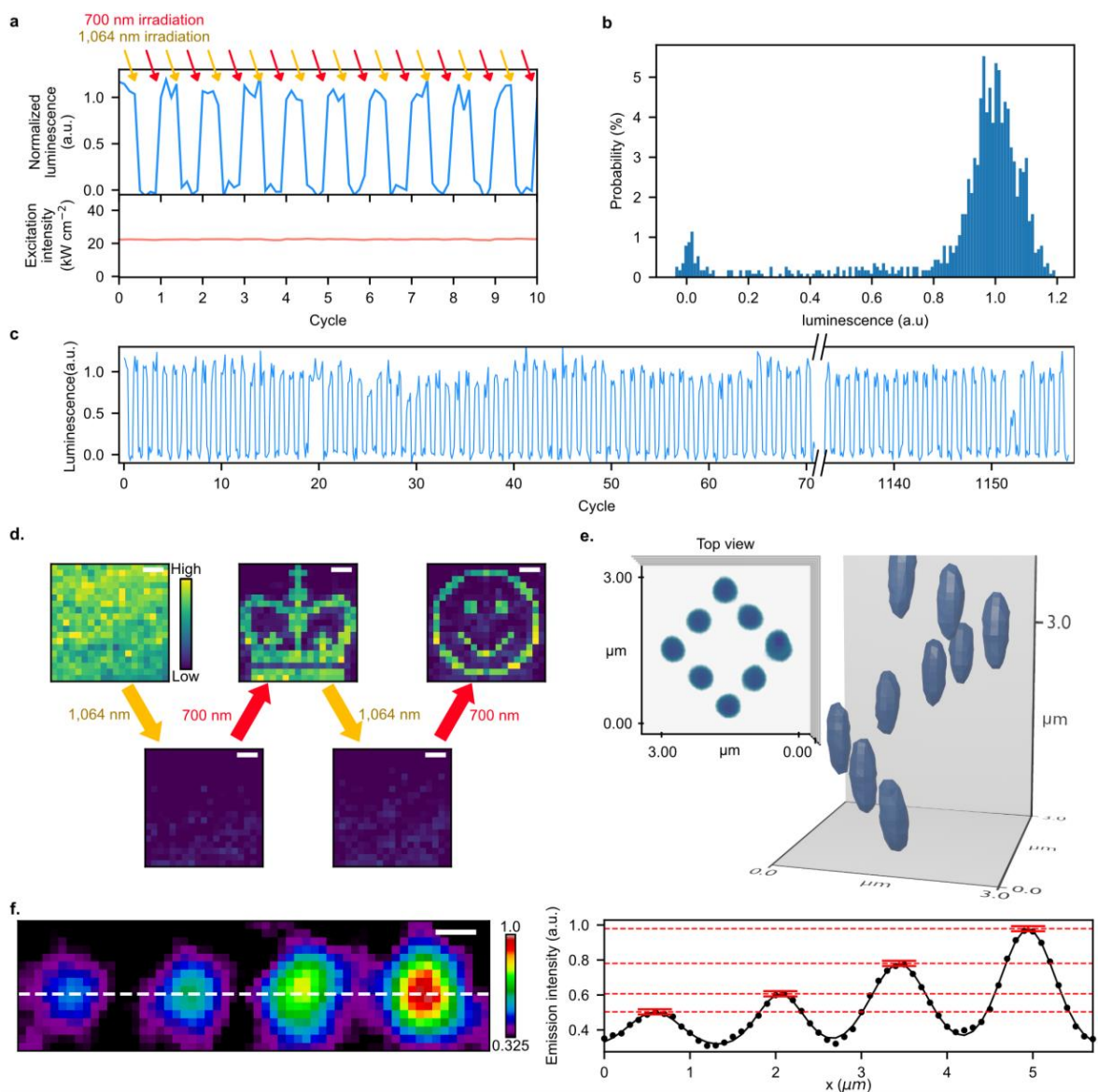

**Fig. 4| 2D and 3D microscale optical write, erase, and rewrite of stable NIR-photoswitchable ANPs. a,** Time-resolved luminescence and excitation intensities for a *single* PEG-coated ANP under ambient conditions. A 1064 nm pump intensity of 22.8 kW cm$^{-2}$ is continuously applied, which excites detectable emission in the bright state but not in the dim state. Irradiation conditions for darkening are 75.5 kW cm$^{-2}$ at 1064 nm and 5 seconds (yellow arrows); and for photobrightening are 164.0 kW cm$^{-2}$ at 700 nm for 10 seconds (red arrows). **b,** A probability histogram of the average emission intensity of the brightened ANP for 1158 irradiation cycles. **c,** Trace of emission from the single PEG-ANP for the first 70 and last 25 irradiation cycles. **d,** Rewritable photopatterning of successive crown and face designs in a 100 nm-thick ANP film. The applicable pixels in the sample were irradiated with intensity of 435 kW cm$^{-2}$ at 1064 nm for 7 s for erasing, and 164 kW cm$^{-2}$ at 700 nm for 5 s for positive lithography. Scale bars are 5 μm. Color bar: normalized luminescence intensity. Side (**e**) and top (**e** inset) views of a 3D rendered image of a diamond spiral optically patterned into a ~5 micron-thick ANP film. The dark voxels in **e** represent emission intensity reduced >70% relative to average emission intensity before photodarkening. **f,** An Image (left) of spots in a photodarkened 8% Tm$^{3+}$ ANP film photobrightened with increasing intensities of a 700 nm focused beam. (right) A linecut from the white dashed line. Scale bar is 500 nm. Color bar: normalized luminescence intensity. Red dashed lines and red error bars are the amplitudes and standard deviations of fitted

emission spectra due to transitions at the $^1I_6$ state in Fig. S15) and ultimately results in the transfer of an electron from the Tm$^{3+}$ ion or the neighboring atoms to the NaYF$_4$ conduction band (Fig. 3d), which is then trapped in a local defect state. Because the process is fully reversible (see below) and does not involve surface quenching, this charge is likely trapped in the interior of the ANP, possibly at the core/shell interface[43], where minor defects may arise during synthesis but are not be visible even by high-resolution TEM (Fig. 2e). As reported previously, such trap states are capable of quenching the ~0.7 eV $^3F_4$ transition of Tm$^{3+}$ via energy transfer[16,17,43,49], which would increase $W_2$ and, critically, shift $I_{th}$ as observed[22] (Figs. 1b, 3a, and S13, Supplementary information). Studies of other NP systems have shown that a single trap site is sufficient to quench luminescence for nanoparticles of similar dimensions as ANPs[46,50]. Several photoionization mechanisms in Tm$^{3+}$-doped glass fibers under NIR excitation have been proposed[10,39,41,51], and detailed identification of all transitions involved in the current system will require further in-depth study.

We further investigated photoswitching behavior by measuring the emission of single 8% Tm$^{3+}$ core/shell ANPs while sequentially exposing them to repeated cycles of 1064 nm photodarkening followed by 700 nm photobrightening in ambient (Fig. 4a,c) and aqueous environments (Fig. S18). Successful photoswitching of a single ANP was observed over 1158 cycles in either ambient or aqueous conditions without any permanent photodegradation (Fig. 4c, Fig. S18). A probability histogram of single ANP photoswitching shows that emission intensities overwhelmingly return to their original values (Fig. 4b). Partial emission recovery is occasionally observed (*e.g.*, cycle 1152 in Fig. 4c), possibly originating from the involvement of additional trap or defect states[16,17,43,49].

To determine if ANP photoswitching can be leveraged in high-density patterning applications[52], we deposited a thick (~5 micron) film of 8% Tm$^{3+}$ ANPs onto a glass slide to darken and brighten in sequential 2- and 3-dimensional (2D and 3D) patterns, which are then imaged by confocal microscopy (Fig. 4d-e). Rewritable 2D patterns (Fig. 4d) were created using continuous-wave (CW) 1064 nm and 700 nm focused beams for imaging/darkening and writing, respectively. The large nonlinearity of PA emission and photodarkening enables patterning resolutions <70 nm[22], with potential for exceptionally long storage lifetimes and unlimited read-write cycles due to ANP photostability (Fig. 4c and Fig. S18). Because the low scattering of NIR wavelengths in this process enable subsurface imaging into samples[47,53], we attempted optical patterning of a complex 3D design ~3 microns across in all 3 dimensions (Fig. 4e). The patterned diamond spiral shows voxel-to-voxel addressability that expands rewritable functionality beyond the existing thermal reset approach[54]. Furthermore, multiple grayscale levels of 800 nm Tm$^{3+}$ emission can be achieved in an ANP film using varying darkening wavelengths and photon doses, so that a single voxel can host 5 (Fig. 4f) or more levels, expanding density by enabling multi-bit storage per voxel.

To determine how indefinite ANP photoswitching affects localization accuracies in superresolution microscopy techniques, we used Indefinite NIR Photon Avalanching Localization Microscopy (IN-PALM) to image single ANPs (Fig. 5a-e) and ANP clusters (Fig. 5d,e, Fig. S19 and Methods). In related SMLM methods[1], accuracies are typically limited by the number of photons collected from an emissive probe before photobleaching[1,55]. As with photoactivated localization microscopy (PALM)[56], in IN-PALM, the concentration of emitting ANPs is actively controlled at a very low level. Only a subset of probes is emissive at a given time, enabling non-overlapping point spread functions of emitted photons to determine precise centroid fittings and localization accuracies. Unlike some SMLM methods where probes are irreversibly photobleached, ANPs may be brightened and darkened repeatedly, greatly expanding the number of collected photons to improve localization accuracies[1]. We note that acquisition rates are mainly determined by photobrightening exposure times, which are power-dependent (Fig. 3c). Thus, a trade-off exists between exposure time and irradiation intensity, which will depend on sample considerations.

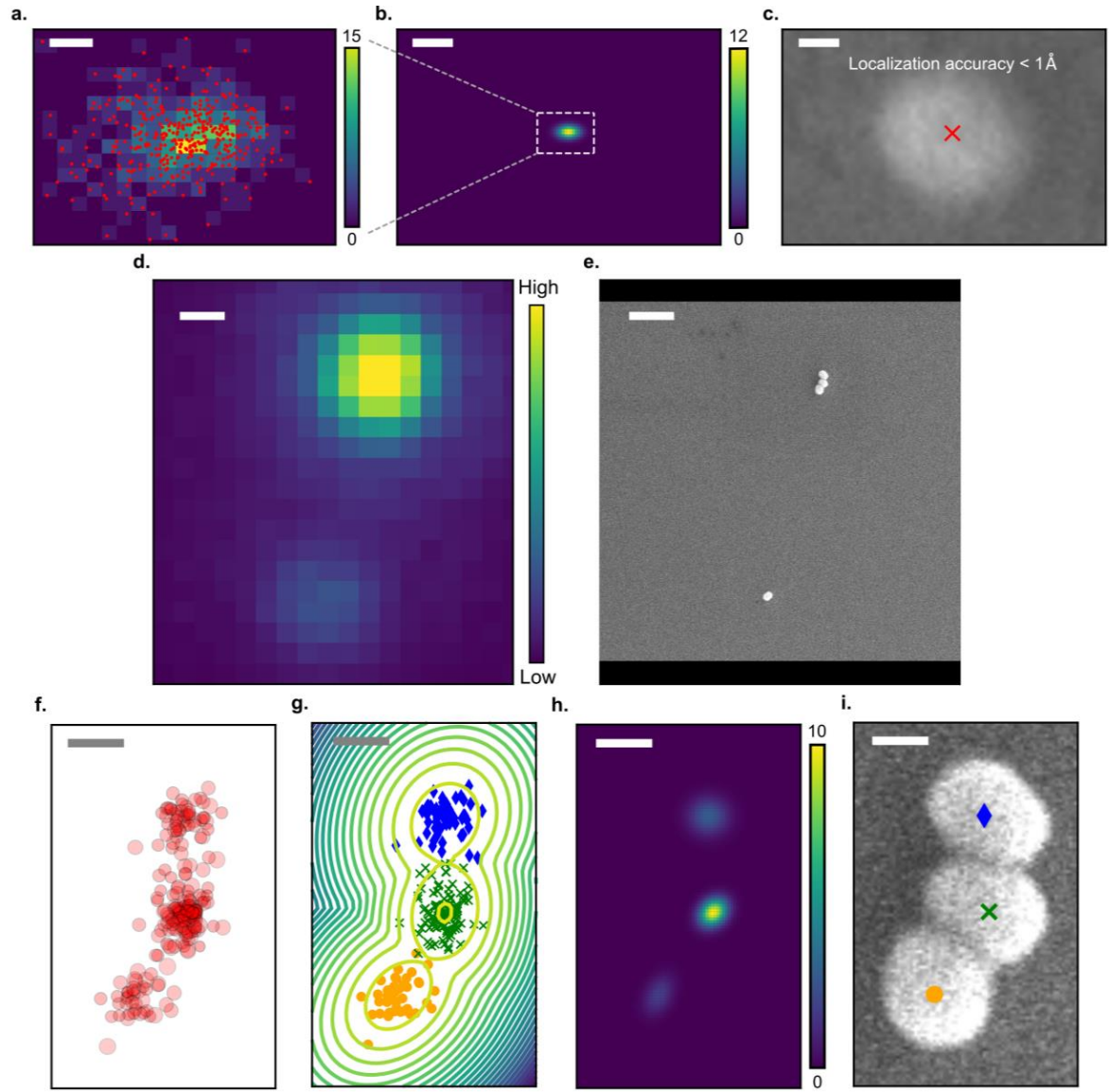

**Fig. 5| Photoactivated localization microscopy of ANPs. a,** 2D histogram of the frame-by-frame localizations of a single 8% $Tm^{3+}$ ANP ($N$ = 403 frames; ANP is photoswitched off and on for each frame). Scale bar is 3 nm. **b,** Fitting of a 2D Gaussian function to the 2D histogram data in **a**. Color bar in **a** and **b**: number of localizations in each pixel. **c,** The centroid derived from the 2D Gaussian fit overlaid on a SEM image of the same ANP. Scale bars in **b** and **c** are 15 nm. Confocal scanning image (**d**) and SEM (**e**) image of a single and a trimer of 8% $Tm^{3+}$ ANPs. Scale bars in **d** and **e** are 200 nm. Color bar in **d**: normalized luminescence intensity. **f,** Frame-by-frame localizations of the trimer in **e** ($N$ = 172). **g,** Clustering of localizations in **f** using Gaussian mixture method. **h,** fitting of a 2D Gaussian to the 2D-histogram data of the clustered localization in **g**. Color bar: number of localizations in each pixel. **i,** The centroids derived from the 2D Gaussian fits overlaid on a SEM image of the trimer. Localization accuracies (symbol, x-axis precision, y-axis precision) are: (blue diamond, 0.52 nm, 0.48 nm), (green cross, 0.33 nm, 0.33 nm), (yellow circle, 0.50 nm, 0.64 nm). Scale bars in **f**, **g**, **h**, and **i** are 20 nm.

Far-field optical and scanning electron microscopy (SEM) images of a region containing a single 26-nm core/4-nm shell ANP and an ANP trimer (Fig. 5d,e) show that individual ANPs in the trimer cannot be resolved optically under wide-field illumination or with confocal scanning (even with the ~70 nm resolution possible due to the extreme nonlinearity of ANPs[22]). Using IN-PALM, we acquired diffraction-limited images of the photoactivated ANPs through repeated photoswitching cycles to reconstruct a superresolved image in which the three touching ANPs are clearly resolved (Fig. 5f-i). The high-precision localization is highlighted (Fig. 5a-c) with a 2D histogram of the localizations from more than 400 IN-PALM frames on a single ANP (Fig 5a; the ANP is photoswitched off and on for each frame). Approximately $1.33 \times 10^8$ total photons were collected from that ANP without signs of degradation, averaging nearly 330,000 per frame. This enables fitting of a 2D Gaussian function to the statistics data[57] (Fig 5b) and calculation of localization accuracies of 0.98 Å and 0.58 Å in the long and short axis, respectively (Fig. 5c). See Methods for additional data processing details.

In conclusion, we report unlimited NIR photoswitching in inorganic ANPs, showing they are photodarkened under NIR-II irradiation and recover with NIR-I or visible irradiation, with no measurable degradation of ANP emission over >1000 repeated photoswitching cycles under both ambient and aqueous conditions. The key mechanism of photoswitching in ANPs is revealed to be a discrete shift of PA threshold intensity. Further, we demonstrate superresolution imaging of ANPs with sub- Å localization precision, as well as rewritable 2D and 3D multi-hued optical patterning with modest NIR lasers. These results open new pathways in a variety of applications including superresolution imaging[4,26], high-density optical memory[1,8,58-60], and robust patterning in both 2 and 3 dimensions[54,61].

*References:*

1   Moerner, W. E. Single-molecule spectroscopy, imaging, and photocontrol: foundations for super-resolution microscopy (nobel lecture). *Angewandte Chemie International Edition* **54**, 8067-8093 (2015).
2   Keller, J., Schönle, A. & Hell, S. W. Efficient fluorescence inhibition patterns for RESOLFT microscopy. *Optics express* **15**, 3361-3371 (2007).
3   Heilemann, M. *et al.* Subdiffraction-resolution fluorescence imaging with conventional fluorescent probes. *Angewandte Chemie International Edition* **47**, 6172-6176 (2008).
4   Dempsey, G. T., Vaughan, J. C., Chen, K. H., Bates, M. & Zhuang, X. Evaluation of fluorophores for optimal performance in localization-based super-resolution imaging. *Nature methods* **8**, 1027-1036 (2011).
5   Hou, L. *et al.* Optically switchable organic light-emitting transistors. *Nature nanotechnology* **14**, 347-353 (2019).
6   Waldermann, F. *et al.* Creating diamond color centers for quantum optical applications. *Diamond and Related Materials* **16**, 1887-1895 (2007).
7   Dhomkar, S., Henshaw, J., Jayakumar, H. & Meriles, C. A. Long-term data storage in diamond. *Science advances* **2**, e1600911 (2016).
8   Gu, M., Zhang, Q. & Lamon, S. Nanomaterials for optical data storage. *Nature Reviews Materials* **1**, 1-14 (2016).
9   Velema, W. A., Szymanski, W. & Feringa, B. L. Photopharmacology: beyond proof of principle. *Journal of the American Chemical Society* **136**, 2178-2191 (2014).
10  Barber, P., Paschotta, R., Tropper, A. & Hanna, D. Infrared-induced photodarkening in Tm-doped fluoride fibers. *Optics letters* **20**, 2195-2197 (1995).

11  Laperle, P., Chandonnet, A. & Vallée, R. Photoinduced absorption in thulium-doped ZBLAN fibers. *Optics letters* **20**, 2484-2486 (1995).
12  Booth, I. J., Archambault, J.-L. & Ventrudo, B. F. Photodegradation of near-infrared-pumped $Tm^{3+}$-doped ZBLAN fiber upconversion lasers. *Optics letters* **21**, 348-350 (1996).
13  Liu, Y. et al. Thermal bleaching of photodarkening and heat-induced loss and spectral broadening in $Tm^{3+}$-doped fibers. *Optics Express* **28**, 21845-21853 (2020).
14  Qin, G. et al. Photodegradation and photocuring in the operation of a blue upconversion fiber laser. *Journal of Applied Physics* **97**, 126108, doi:10.1063/1.1943513 (2005).
15  Xing, Y.-b. et al. Active radiation hardening of Tm-doped silica fiber based on pump bleaching. *Optics express* **23**, 24236-24245 (2015).
16  Hu, Y. et al. X-ray-Excited Super-Long Green Persistent Luminescence from $Tb^{3+}$ Monodoped β-$NaYF_4$. *The Journal of Physical Chemistry C* **124**, 24940-24948 (2020).
17  Ou, X. et al. High-resolution X-ray luminescence extension imaging. *Nature* **590**, 410-415 (2021).
18  Qin, G. et al.   (American Institute of Physics, 2005).
19  Chan, E. M., Levy, E. S. & Cohen, B. E. Rationally designed energy transfer in upconverting nanoparticles. *Advanced Materials* **27**, 5753-5761 (2015).
20  Wu, S. et al. Non-blinking and photostable upconverted luminescence from single lanthanide-doped nanocrystals. *Proceedings of the National Academy of Sciences* **106**, 10917-10921 (2009).
21  Wen, S. et al. Future and challenges for hybrid upconversion nanosystems. *Nature Photonics* **13**, 828-838 (2019).
22  Lee, C. et al. Giant nonlinear optical responses from photon-avalanching nanoparticles. *Nature* **589**, 230-235 (2021).
23  Müller, T., Schumann, C. & Kraegeloh, A. STED microscopy and its applications: new insights into cellular processes on the nanoscale. *ChemPhysChem* **13**, 1986-2000 (2012).
24  Lukinavičius, G. et al. A near-infrared fluorophore for live-cell super-resolution microscopy of cellular proteins. *Nature chemistry* **5**, 132-139 (2013).
25  Dempsey, G. T., Vaughan, J. C., Chen, K. H., Bates, M. & Zhuang, X. Evaluation of fluorophores for optimal performance in localization-based super-resolution imaging. *Nature methods* **8**, 1027 (2011).
26  Chozinski, T. J., Gagnon, L. A. & Vaughan, J. C. Twinkle, twinkle little star: photoswitchable fluorophores for super-resolution imaging. *FEBS letters* **588**, 3603-3612 (2014).
27  Berkovic, G., Krongauz, V. & Weiss, V. Spiropyrans and spirooxazines for memories and switches. *Chemical reviews* **100**, 1741-1754 (2000).
28  Irie, M. Diarylethenes for memories and switches. *Chemical Reviews* **100**, 1685-1716 (2000).
29  Beharry, A. A. & Woolley, G. A. Azobenzene photoswitches for biomolecules. *Chemical Society Reviews* **40**, 4422-4437 (2011).
30  Brakemann, T. et al. A reversibly photoswitchable GFP-like protein with fluorescence excitation decoupled from switching. *Nature biotechnology* **29**, 942-947 (2011).
31  Dickson, R. M., Cubitt, A. B., Tsien, R. Y. & Moerner, W. E. On/off blinking and switching behaviour of single molecules of green fluorescent protein. *Nature* **388**, 355-358 (1997).
32  Erno, Z., Yildiz, I., Gorodetsky, B., Raymo, F. M. & Branda, N. R. Optical control of quantum dot luminescence via photoisomerization of a surface-coordinated, cationic dithienylethene. *Photochemical & Photobiological Sciences* **9**, 249-253 (2010).
33  Park, Y. I. et al. Nonblinking and nonbleaching upconverting nanoparticles as an optical imaging nanoprobe and T1 magnetic resonance imaging contrast agent. *Advanced Materials* **21**, 4467-4471 (2009).
34  Nam, S. H. et al. Long-term real-time tracking of lanthanide ion doped upconverting nanoparticles in living cells. *Angewandte Chemie International Edition* **50**, 6093-6097 (2011).

35  Gargas, D. J. *et al.* Engineering bright sub-10-nm upconverting nanocrystals for single-molecule imaging. *Nature nanotechnology* **9**, 300-305 (2014).
36  Fernandez-Bravo, A. *et al.* Continuous-wave upconverting nanoparticle microlasers. *Nature nanotechnology* **13**, 572-577 (2018).
37  Fernandez-Bravo, A. *et al.* Ultralow-threshold, continuous-wave upconverting lasing from subwavelength plasmons. *Nature materials* **18**, 1172-1176 (2019).
38  Akhmetov, S., Akhmetova, G., Kolodiev, B. & Samojlovich, M. Optical absorbtion spectra of YAG ($Eu^{2+}$, $Eu^{3+}$) and YAG ($Yb^{2+}$, $Yb^{3+}$) crystals. *Zhurnal Prikladnoj Spektroskopii* **48**, 681-683 (1988).
39  Broer, M., Krol, D. & DiGiovanni, D. Highly nonlinear near-resonant photodarkening in a thulium-doped aluminosilicate glass fiber. *Optics letters* **18**, 799-801 (1993).
40  Engholm, M. & Norin, L. Preventing photodarkening in ytterbium-doped high power fiber lasers; correlation to the UV-transparency of the core glass. *Optics Express* **16**, 1260-1268 (2008).
41  Lupi, J.-F. *et al.* Steady photodarkening of thulium alumino-silicate fibers pumped at 1.07 μm: quantitative effect of lanthanum, cerium, and thulium. *Optics letters* **41**, 2771-2774 (2016).
42  Zhuang, Y. *et al.* X-ray-charged bright persistent luminescence in $NaYF_4$: $Ln^{3+}$@ $NaYF_4$ nanoparticles for multidimensional optical information storage. *Light: Science & Applications* **10**, 1-10 (2021).
43  Qin, X. *et al.* Suppression of Defect-Induced Quenching via Chemical Potential Tuning: A Theoretical Solution for Enhancing Lanthanide Luminescence. *The Journal of Physical Chemistry C* **123**, 11151-11161 (2019).
44  Liang, Y. *et al.* Migrating photon avalanche in different emitters at the nanoscale enables 46th-order optical nonlinearity. *Nature Nanotechnology*, 1-7 (2022).
45  Bednarkiewicz, A., Chan, E. M., Kotulska, A., Marciniak, L. & Prorok, K. Photon avalanche in lanthanide doped nanoparticles for biomedical applications: super-resolution imaging. *Nanoscale Horizons* (2019).
46  Kwock, K. W. *et al.* Surface-Sensitive Photon Avalanche Behavior Revealed by Single-Avalanching-Nanoparticle Imaging. *The Journal of Physical Chemistry C* **125**, 23976-23982 (2021).
47  Levy, E. S. *et al.* Energy-looping nanoparticles: harnessing excited-state absorption for deep-tissue imaging. *ACS nano* **10**, 8423-8433 (2016).
48  Qin, X., Liu, X., Huang, W., Bettinelli, M. & Liu, X. Lanthanide-activated phosphors based on 4f-5d optical transitions: theoretical and experimental aspects. *Chemical reviews* **117**, 4488-4527 (2017).
49  Narasimha Reddy, K. & Subba Rao, U. High-temperature X-ray irradiation induced thermoluminescence and half-life calculations in $NaYF_4$ polycrystalline samples. *Crystal Research and Technology* **19**, 1399-1403 (1984).
50  Zhang, Y., Lei, P., Zhu, X. & Zhang, Y. Full shell coating or cation exchange enhances luminescence. *Nature communications* **12**, 1-10 (2021).
51  Chandonnet, A., Laperle, P., LaRochelle, S. & Vallée, R. in *Photosensitive Optical Materials and Devices.* 70-81 (SPIE).
52  Ren, Y. *et al.* Reversible upconversion luminescence modification based on photochromism in $BaMgSiO_4$: $Yb^{3+}$, $Tb^{3+}$ ceramics for anti-counterfeiting applications. *Advanced Optical Materials* **7**, 1900213 (2019).
53  Tian, B. *et al.* Low irradiance multiphoton imaging with alloyed lanthanide nanocrystals. *Nature communications* **9**, 1-8 (2018).
54  Hu, Z. *et al.* Reversible 3D optical data storage and information encryption in photo-modulated transparent glass medium. *Light: Science & Applications* **10**, 1-9 (2021).

55	Thompson, M. A., Lew, M. D. & Moerner, W. Extending microscopic resolution with single-molecule imaging and active control. *Annual review of biophysics* **41**, 321-342 (2012).
56	Betzig, E. *et al.* Imaging intracellular fluorescent proteins at nanometer resolution. *science* **313**, 1642-1645 (2006).
57	Cnossen, J. *et al.* Localization microscopy at doubled precision with patterned illumination. *Nature methods* **17**, 59-63 (2020).
58	Xie, N.-H. *et al.* Deciphering erasing/writing/reading of near-infrared fluorophore for nonvolatile optical memory. *ACS applied materials & interfaces* **11**, 23750-23756 (2019).
59	Adam, V. *et al.* Data storage based on photochromic and photoconvertible fluorescent proteins. *Journal of biotechnology* **149**, 289-298 (2010).
60	Padgaonkar, S. *et al.* Light-Triggered Switching of Quantum Dot Photoluminescence through Excited-State Electron Transfer to Surface-Bound Photochromic Molecules. *Nano letters* **21**, 854-860 (2021).
61	Zhuang, Y., Wang, L., Lv, Y., Zhou, T. L. & Xie, R. J. Optical data storage and multicolor emission readout on flexible films using deep-trap persistent luminescence materials. *Advanced Functional Materials* **28**, 1705769 (2018).

# Methods

## Materials

Sodium trifluoroacetate (Na-TFA, 98%), sodium oleate, ammonium fluoride ($NH_4F$), yttrium chloride ($YCl_3$, anhydrous, 99.99%), thulium chloride ($TmCl_3$, anhydrous, 99.9+%), gadolinium chloride ($GdCl_3$, anhydrous, 99.99%), yttrium trifluoroacetate (99.99+%), oleic acid (OA, 90%), and 1-octadecene (ODE, 90%) were purchased from Sigma-Aldrich.

## Synthesis of core and core/shell ANPs

$NaY_{1-x}Tm_xF_4$ ANP cores with average diameters ranging from d = 10 to 18 ± 1 nm (see Table S1) were synthesized based on reported procedures[47,53]. For x = 0.01 (i.e., 1% $Tm^{3+}$ doping), $TmCl_3$ (0.01 mmol, 2.8 mg) and $YCl_3$ (0.99 mmol, 193.3 mg) and were added into a 50 ml 3-neck flask following injection of 6 ml of OA and 14 ml of ODE. The mixture was stirred under vacuum and heated to 100 °C for 1 h. The solution was pumped with vacuum and purged with $N_2$ over three cycles to remove water and oxygen. Subsequently, $NH_4F$ (4 mmol, 148 mg) and sodium oleate (2.5 mmol, 762 mg) were added to the flask with $N_2$ gas flow. Afterward, the flask was resealed and placed under vacuum for 15 min at 100 °C, followed by analogous 3 cycles of alternating vacuum pump and $N_2$ purge for additional 10 min. After that, the solution was quickly heated to 320 °C (the approximate ramp rate: 25 °C/min). The temperature stayed at 320 °C for 40 - 60 min. The solution was cooled to room temperature with compressed air.

Ethanol was added to a tube containing the ANPs and the nanocrystals were separated by centrifugation for 5 min at 4000 rpm. The dispersion with suspended pellets was additionally centrifuged to remove large aggregated particles. The nanoparticles were purified by a combination of ethanol wash, centrifuging, and pellet dissolution in hexane. The whole cycle is repeated one more time to further purify the nanocrystals. The nanocrystals were stored in hexane with two drops of OA to prevent aggregation.

## Shell growth

A 0.1 M stock solution of 20% $GdCl_3$ and 80% $YCl_3$ was prepared by mixing $YCl_3$ (2 mmol, 390.5 mg), $GdCl_3$ (0.5 mmol, 131.8 mg), 10 ml OA and 15 ml ODE in a 50 ml 3-neck flask. The mixture was stirred under vacuum and heated to 110 °C for 30 min. The flask was filled with $N_2$ gas and heated to 200 °C for about 1 h, until the solution became clear, and no solid was seen in the solution. The solution was cooled to 100 °C and placed under vacuum for 30 min. A 0.2 M solution of Na-TFA was prepared by mixing Na-TFA (4 mmol, 544 mg), 10 ml OA and 10 ml ODE in a flask, under vacuum, at room temperature for 2 h to ensure that all chemicals were dissolved. 3-9 nm $NaY_{0.8}Gd_{0.2}F_4$ shells (see Table S1) were grown on ANP cores using a layer-by-layer protocol[62] inside a nitrogen-filled glovebox containing Workstation for Automated Nanocrystal Discovery and Analysis (WANDA)[62]. For example, for a 3 nm shell thickness, 6 mL ODE and 4 mL OA were injected to the dried ANP cores and heated to 280 °C at 20 °C/min in the WANDA glove box. a 0.2 M Na-TFA stock solution and a 0.1 M stock solution of 20% Gd and 80% Y oleate solution was added alternatively according to the automated protocols. Each alternating injection cycle was performed every 40 minutes (e.g., one injection every 20 minutes) over 6 repeated cycles. After the last injection of each cycle, it was annealed at 280 °C for an additional 30 minutes. After that, it was cooled rapidly by $N_2$ gas flow. The core-shell particles were separated and purified using an identical purification protocol described above.

Core-shell $NaYF_4$ nanocrystals with varying $Tm^{3+}$ concentrations (from 1 to 100%) were fabricated using analogous protocols.

**Nanoparticle characterization**
Transmission electron microscopy (TEM) was achieved using a JEOL JEM-2100F field emission TEM operating at an acceleration voltage of 200 kV, FEI Themis 60-300 STEM/TEM at an acceleration voltage of 300 kV and Tecnai T20 S-TWIN TEM at 200 kV with a LaB6 filament. The statistics of the nanocrystal sizes were calculated based on the size of approximately 100 nanoparticles using ImageJ software. X-Ray diffraction (XRD) measurement was performed using a Bruker D8 Discover diffractometer with Cu Kα radiation.

The high-resolution STEM images were acquired on an aberration corrected Titan 80-300 called TEAM 0.5 at the Molecular Foundry. The microscope was operated at 200 kV with a convergence semi-angle of 17 mrad and approximately 5 pA beam current. The 4D Camera was used to acquire a series of diffraction patterns from a grid of 1024x1024 probe positions with real space step size of 40 pm and acquisition time of 11 microseconds. The center of mass of each diffraction pattern was calculated and then used to estimate the phase of the electron beam by the differential phase contrast technique. Each algorithm was implemented in the open source stempy package. The phase is much more sensitive to weakly scattering atoms such as Fluorine compared to the Z-contrast of the annular dark field signal. This allowed us to image the atomic structure of beam sensitive ANPs to confirm they do not contain large scale defects.

**Preparation of nanocrystal film samples**
To prepare film samples, nanocrystals in a 40 µL suspension with a concentration of 1 µM were either drop-casted or spin-coated on a coverslip. AFM measurements (Bruker Dimension AFM) were performed to measure the film thickness of the prepared films.

**Optical characterization of ANPs**
For confocal microscopy of ANPs, an inverted confocal microscope (Nikon, Eclipse Ti-S) fitted with a 3D (XYZ) nanoscanning piezo stage (Physik Instrumente, P-545.xR8S Plano) was used. Single particles deposited on glass coverslips were excited with a 1064-nm continuous-wave diode laser (Thorlabs, M9-A64-0300). A 950 nm long-pass filter (Thorlabs, FELH 950) and 950 nm short-pass dichroic mirror (Thorlabs, DMSP 950) were placed on the excitation beam path to filter out all the wavelengths above 950 nm. An 850 nm short-pass filter (Thorlabs, FESH 850) and a 750 nm long-pass filter (Thorlabs, FELH 750) were used to selectively collect the 800 nm photons from the sample. A 1.49NA 100× immersion oil objective (Olympus) and a 0.95NA 100X air objective lens (Nikon) were used for the imaging of single ANPs and ANP ensembles, respectively. Emitted light was directed to an electron-multiplying charge-coupled device (EMCCD)-equipped spectrometer (Princeton Instruments, ProEM: $1600^2$ eXcelon™3) or a single-photon avalanche diode (Micro Photon Device, PDM series). A neutral density wheel with a continuously variable density (Newport, 100FS04DV.4) was synchronized with the collection system and automatically rotated by an Arduino-controlled rotator to perform power dependence measurements. A Thorlabs power meter (PM100D and S120VC) simultaneously recorded the ~10% of the laser power reflected by a glass coverslip. Average excitation power densities were estimated using measured laser powers on the sample plane converted by the recorded laser power by the power meter and using the area calculated from the FWHM of the imaged laser spot.

**Time-Tagged Time-Resolved (TTTR) luminescence**
For time-resolved luminescence experiment, a time-correlated single-photon counting (TCSPC) device (Picoquant, Hydraharp 400) coupled to a single-photon avalanche diode was used to record the timing data of detected photons. Time-Tagged Time-Resolved (TTTR) luminescence of ANP ensembles was measured by detecting single photons and recording the arrival time relative to the beginning of the measurement.

**Photoswitching of ANP ensemble films with two lasers**
To investigate the photoswitching properties of ANP ensembles, illumination at 1064 nm and 510-950 nm from a 1064-nm continuous-wave diode laser (Thorlabs, M9-A64-0300), and a Ti-sapphire pulsed laser (Coherent, Chameleon OPO Vis, 80 MHz) or a 532 nm continuous-wave laser (Coherent, Sapphire CDRH) were focused on the ANPs on the inverted confocal microscope. A 950 nm short-pass dichroic mirror (Thorlabs, DMSP950R) was placed in the excitation beam path to merge two laser beams for photodarkening and photobrightening. The alignment of two laser beams at the sample plane was confirmed by measuring the beam images on a CMOS camera (Amscope, MU503) installed on a side port of the inverted microscope. The timing of the two illuminations was programmed by Scopefoundry, a custom python-based software, which controls the two dual-position sliders (Thorlabs, ELL6).

**Correlative light and electron microscopy for indefinite NIR photon avalanching localization microscopy (IN-PALM)**
The single ANPs on a glass coverslip marked by a finder grid (Gilder Grids, G200F1-C3) were placed on an inverted microscope (Nikon, Eclipse Ti2000-U). Wide-field illumination from a 1064-nm continuous-wave diode laser (3SP Technologies, 1064CHP) and focused illumination from a 532 nm continuous-wave laser (Cobolt, Samba 50 mW) were directed on the sample through a 1.4NA 50× immersion oil objective (Nikon, PLAN APO 60x). The two-color illuminations were alternated using two stepper motors which open and close a beam block on the signal of an Arduino board controlled by a python-based program (Scopefoundry). To shift from 1064 nm illumination for imaging to that for photodarkening, the beam size of the wide-field illumination was changed using a motorized flipper (Newport, 8892-K-M) in which a plano-convex lens (Thorlabs, AC254-400) is mounted. The 1064 nm excitation intensities for imaging and photodarkening are 58.7 kW cm-2 and 559 kW cm-2, respectively. The 532 nm excitation intensity for photobrightening is 842 kW cm-2. The probability of the photobrightening occurrence is governed by the statistical distribution in Fig. S24. The exposure times for darkening and brightening are set to 2 and 8 sec, respectively. The samples under 1064 nm excitation were imaged by an EMCCD camera (Andor, iXon DU-888D-C00-#BV). The EM gain and the exposure time of the camera were set to 300 and 1 sec, respectively. The scanning electron microscopy was performed using 10 kV SEM (Carl Zeiss, SigmaHD) after the sample coated with platinum for 120 s using a sputter coater (Cressington, 108).

**Data processing for IN-PALM**
To reconstruct and quantify IN-PALM images, the centroid of PSFs was estimated using a 2D-gaussian fitting. The fitting of 2D-gaussian function to the wide-field image frames offers the collection of the PSF centroids dispersed along with the drift of the sample (Fig. S20c and S21). The drift correction was achieved by using reference particle images which are not photoswitched during IN-PALM imaging. The PSFs of the 6 ANPs collected during the fitting was used to correct the lateral and rotational drift of the sample, yielding centroid clusters within 20 nm (Fig. S22). The 2D histograms of the centroid clusters (Fig. 5a;

ANP is photoswitched off and on for each frame) allows for the further fitting of 2D-gaussian function to the statistic images (Fig. 5b) which provides the centroid positions and the localization accuracies of the ANP clusters.

Data-processing using a Gaussian mixture method offers a straight-forward way to separate a mixture of localizations into several groups (Fig. 5f, g)[63]. Also, the near-uniform brightness of each ANP proves beneficial, allowing the use of a simple intensity filter to reject frames that include more than one brightened ANP within a cluster, eliminating another source of error (Fig. S23). In rare instances, more than one particle can partially photobrighten, resulting in erroneous position localization estimates.

**Methods References:**

62   Chan, E. M. *et al.* Combinatorial discovery of lanthanide-doped nanocrystals with spectrally pure upconverted emission. *Nano letters* **12**, 3839-3845 (2012).
63   Lemmer, M., Inkpen, M. S., Kornysheva, K., Long, N. J. & Albrecht, T. Unsupervised vector-based classification of single-molecule charge transport data. *Nature communications* **7**, 1-10 (2016).

**Data availability**
All data generated or analysed during this study, which support the plots within this paper and other findings of this study, are included in this published article and its Supplementary Information. Source data are provided with this paper.

**Code availability**
The code for modelling the PA behaviour using the differential rate equations described in the Supplementary Information are freely available at https://github.com/nawhgnahc/Photon_Avalanche_DRE_calculation.git.

**Acknowledgements:**
P.J.S., Y.D.S., S.H.N., J.K. and C.L. gratefully acknowledge support from the Global Research Laboratory (GRL) Program through the National Research Foundation of Korea (NRF) funded by the Ministry of Science and ICT (number 2016911815), and KRICT (KK2261-12, SKO1930-20). E.Z.X. gratefully acknowledges support from the NSF Graduate Research Fellowship Program. K.W.C.K. acknowledges support from the DOE NNSA Laboratory Residency Graduate Fellowship program (No. DE-NA0003960). Work at the Molecular Foundry was supported by the Office of Science, Office of Basic Energy Sciences, of the US Department of Energy under contract number DE-AC02-05CH11231. P.J.S. also acknowledges support from Programmable Quantum Materials, an Energy Frontier Research Center funded by the US Department of Energy (DOE), Office of Science, Basic Energy Sciences (BES), under award DE-SC0019443. N.F.M. acknowledges support from the European Union's Horizon 2020 research and innovation program under the Marie Sklodowska-Curie grant agreement No 893439, the Fulbright Scholarship Program, the Zuckerman-CHE STEM Leadership Program, and the ISEF Foundation.

**Author contributions:** PJS, CL, EMC, BEC, and YDS conceived of the study. Experimental measurements and associated analyses were conducted by CL, EZX, KWCK, NFM, CCSP, HSP, JK, SHN, SDP, TL, PE, and EMC. Advanced nanoparticle synthesis and characterization was performed by YL, AT, CCSP, HSP, BEC and EMC. Theoretical modelling and simulations of photon avalanching photophysics were carried out by CL and EMC. All authors contributed to the preparation of the manuscript.

**Competing interests:** the authors declare no competing interests